\documentclass[twocolumn]{revtex4}
\usepackage{graphicx}
\usepackage{dcolumn}
\usepackage{bm}

\begin{document}
\title{Kilohertz laser ablation for doping helium nanodroplets}

\author{M. Mudrich} \email{mudrich@physik.uni-freiburg.de}
    \affiliation{Physikalisches Institut, Universit\"at Freiburg, D-79104 Freiburg, Germany.}
 \author{B. Forkl}
 \affiliation{Physikalisches Institut, Universit\"at Freiburg, D-79104 Freiburg, Germany.}
\author{S. M\"uller}
    \affiliation{Physikalisches Institut, Universit\"at Freiburg, D-79104 Freiburg, Germany.}
\author{M. Dvorak}
    \affiliation{Physikalisches Institut, Universit\"at Freiburg, D-79104 Freiburg, Germany.}
\author{O. B\"unermann}
    \affiliation{Physikalisches Institut, Universit\"at Freiburg, D-79104 Freiburg, Germany.}
\author{F. Stienkemeier}
    \affiliation{Physikalisches Institut, Universit\"at Freiburg, D-79104 Freiburg, Germany.}

\date{\today}

\begin{abstract}
A new setup for doping helium nanodroplets by means of laser ablation at kilohertz
repetition rate is presented. The doping process is characterized and two distinct
regimes of laser ablation are identified. The setup is shown to be efficient and stable
enough to be used for spectroscopy, as demonstrated on beam-depletion spectra
of lithium atoms attached to helium nanodroplets. For the first time, helium droplets are
doped with high temperature refractory materials such as titanium and tantalum. Doping
with the non-volatile DNA basis Guanine is found to be efficient and a number of
oligomers are detected.
\end{abstract}

\maketitle

\section{\label{sec:Intro}Introduction}
Helium nanodroplet isolation (HENDI) has established as a versatile technique for
spectroscopic studies of a variety of atomic and molecular systems, ranging from
microwave and infrared spectroscopy of small molecules to electronic spectroscopy of
molecular complexes~\cite{Toennies:2004}. In particular the property of helium droplets
to efficiently cool dopant particles down to 0.4 K equilibrium temperature and to act as
nanoscale reactors for assembling molecular complexes of virtually any constituents has
stimulated studies of rather unconventional molecular
systems~\cite{Grebenev:2001,Lugovoi:2000}. Recent directions include larger metal
clusters \cite{Tiggesbaeumker:2007} and studies of the dynamics of doped helium droplets
\cite{Stienkemeier:2006}.

Up to now, the choice of dopants has mostly been constrained by the limited possibilities
to bring molecules into the gas phase by thermal evaporation inside a heated cell.
Therefore, only a restricted number of metals and larger organic molecules have found
their way into helium droplet machines~\cite{Stienkemeier:2001}. Using a pyrolysis
source, the range of dopant molecules has been extended to radicals~\cite{Kuepper:2002}.
However, the study of metal clusters including elements such as Fe, Ni, Co, Ti, V, Nb,
Au, Pt is of considerable interest for their particular magnetic or superconducting
properties at very low temperatures and for their catalytic properties. Furthermore, the
ability of doping fragile organic molecules would extend the study of spectra of
biomolecules at 0.4\,K and the optic properties of organic aggregates, the latter being of
particular interest in the context of organic opto-electronic devices. Moreover, doping
helium droplets with ions, as demonstrated using laser ablation~\cite{Claas:2003}, opens
up a new class of experiments. Besides the spectroscopic interest in ions inside doped
helium droplets~\cite{Galli:2001}, ion doped droplets will enable experiments with
size-selected helium droplets.

Different variants of laser ablation sources have been developed for different
applications. Generally, two regimes of interaction of laser radiation with solid
material can be distinguished. In the regime of intermediate power densities, laser
desorption/vaporization leads to the removal of neutral molecules from the surface. The
surface may consist of bulk sample material, of a monolayer of transparent sample on a
strongly absorbing metal substrate~\cite{Posthumus:1978} or of a thick, strongly
absorbing matrix in which a small amount of sample material is doped~\cite{Karas:1987}. Laser
desorption/vaporization is of relevance for vaporizing large, fragile molecules such as
proteins and DNA for gas-phase analysis~\cite{Levis:1994}. In the regime of high laser
power densities, non-thermal processes lead to ejection or ablation of material into the
gas-phase involving complex laser-plasma interaction processes. The latter regime is
particularly interesting as a source of inorganic and metallic
clusters~\cite{Powers:1983,Laihing:1987,Milani:1990,Heiz:1997,Wagner:1997,Haberland:1994} and of free electrons and
ions~\cite{Ghazarian:2002,Claas:2003}. 

The combination of laser desorption with supersonic jets is a widely used technique for
several reasons. Laser desorbed molecules are translationally and internally very hot and
are likely to fragment unless they are cooled by thermalization with cold carrier gas in
the expansion~\cite{Elam:1998}. Furthermore, the molecules are formed in sufficient
abundance and with sufficient shot-to-shot stability to do spectroscopy.

The combination of laser ablation/desorption with a helium droplet beam source has been
reported before only by Ghazarian and co-workers and by our
group~\cite{Claas:2003,Ghazarian:2002}. However, in the experiment of Ref.~\cite{Ghazarian:2002}
droplet formation occurs by postexpansion condensation in a high density helium atmosphere, 
which is quite different compared to a nanodroplet beam traveling under high vacuum conditions~\cite{Claas:2003}.
As with standard molecular beam machines, the
laser ablation target is placed next to the jet expansion to benefit from the high
thermalization rate of ablated particles with high density helium gas. In a second stage,
the atoms or molecules picked up by the helium droplets are further cooled to the
equilibrium temperature of 0.4\,K through evaporation of helium atoms off the droplets.

The motivation for setting up a stable laser-ablation source with kilohertz repetition
rate is manifold. Since laser-ablation sources are not inherently very stable, extensive
signal averaging is generally required. On the one hand, many applications such as
coincidence measurements require small signal rates per laser shot. The only way of
reducing the measuring time is to increase the repetition rate of the experimental
cycles. On the other hand, in applications in which large particle fluxes are needed such
as mass-selected deposition on surfaces, upscaling the repetition rate increases the
efficiency which saves time or allows to study species of low abundance. Furthermore, modern
nanosecond and femtosecond laser systems are typically operated at 1\,kHz or higher
repetition rate. Ideally, the laser ablation source should match the repetition rate of
the laser system. The laser-ablation source presented in this article has been developed to
eventually combine it with a fs-laser system and a pulsed beam of helium nanodroplets,
each being operated at 1\,kHz repetition rate.

In this paper we demonstrate the stable operation of a laser ablation source at kilohertz
repetition rate for doping a beam of helium nanodroplets (He$_N$, $N\sim 10^4$) with
refractory materials and biomolecules. Doping helium droplets with ions using laser
ablation has been demonstrated before~\cite{Claas:2003}. In the following, the design of
the simple yet versatile mechanical setup is detailed. A load-lock system is implemented
for exchanging samples without breaking the vacuum. This is crucial when operating a low
temperature droplet source where venting costs considerable time, generally a day. The
process of doping helium nanodroplets by laser ablation is characterized using lithium
(Li) and magnesium (Mg) as sample materials. Long-term stability is shown to be high
enough to perform laser spectroscopy of laser ablated Li atoms. The applicability of the
new laser ablation source to doping with high temperature refractory metals and with the
non-volatile DNA base guanine is demonstrated.

\section{\label{sec:Experiment}Experiment}
The laser ablation setup is incorporated into the source chamber of a helium droplet
machine, which is described in detail elsewhere~\cite{Claas:2003,Stienkemeier:1999}. In
short, helium gas (stagnation pressure 30-60\,bar) expands through a cold (14-20\,K)
nozzle 10\,$\mu$m in diameter. Behind the skimmer (diameter: 400\,$\mu$m, distance:
16\,mm) the droplet beam passes through a heated cell for doping with Li atoms in order
to compare with doping by laser ablation. Further downstream the doped droplet beam is
analysed using a commercial quadrupole mass spectrometer (QMS) with an electron impact
ionizer (Extrel, mass range 1-1000\,amu). Furthermore, a Langmuir-Taylor surface ionization detector terminates the doped
droplet beam. This detector can accurately measure the number of alkali and alkaline
earth doped helium droplets~\cite{Stienkemeier:2000}.

The advantages of implementing the laser ablation setup directly into the source chamber
are the following: First, this compact geometry allows to produce a beam of doped helium
droplets in only one vacuum chamber. Second, laser ablated atoms and molecules carry a
large amount of kinetic and internal energy. They are efficiently ``precooled'' by
scattering with the relatively high density of helium atoms in the outer parts of beam.
In this way, internal and kinetic energies are sufficiently reduced for the particles to
attach to helium droplets without complete evaporation of the droplets. Furthermore, the
laser ablated particles can directly serve as seeds and enhance droplet condensation.
This effect has been observed in the case of doping with atomic ions~\cite{Claas:2003}.

\begin{figure}
\resizebox{1.0\columnwidth}{!}{\includegraphics{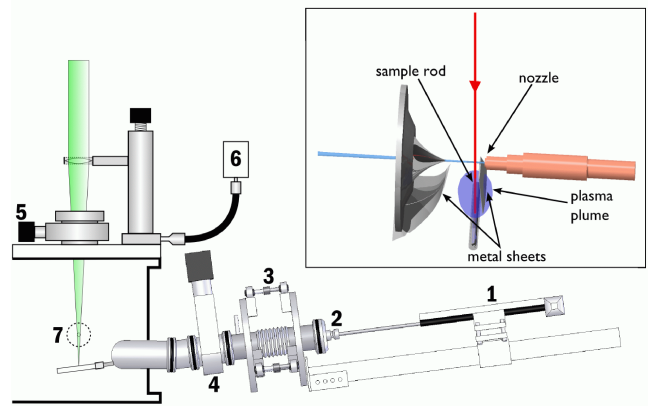}} \caption{\label{fig:setup}
Side view of the mechanical setup for implementing kHz laser ablation into the helium
droplet machine. The ablation laser beam enters the nozzle chamber from above through an
entrance window mounted on a gate valve (5) and is focused onto the sample rod. The helium
droplet beam propagates normal to the paper plane through the skimmer (dashed circle, (7))
right above the target rod. The main technical features are (1) a translation stage mounted upside down,
(2) a rotary feedthrough, (3) a bellow chamber seving as vacuum lock and for aligning the sample, (4, 5) gate valves, and (6) a motor coupled to a translation stage which moves the
focusing lens back and forth. The inset shows the metal sheets that cover the direct line
of sight from the ablation spot to nozzle and skimmer orifices.
}
\end{figure}

The laser ablation setup is depicted in Fig.~\ref{fig:setup}. A target rod (3-10\,mm in
diameter, up to 7\,cm length) is placed $\approx 2\,$cm below the helium beam in between
the cryogenic nozzle and skimmer perpendicularly to the droplet beam. The vertical
position corresponds to a distance of the sample from the droplet beam causing negligible
interferences with the beam expansion. However, moving the vertical position of the
sample in the range of $\pm 10\,$mm has only little influence on the laser doping
efficiency. A focused laser beam enters the source chamber from above through a window
and creates an ablation plume on the target surface which is directed towards the helium
beam. In order to homogeneously ablate the entire rod surface the sample rod is
simultaneously translated and rotated along its axis in a helical fashion. In this way, a
new sample spot is provided to each laser pulse in order to minimize the formation of
grooves. This is realized by coupling the rod to the spindle of a commercial motorized
translation stage mounted upside down onto its moving stage outside the vacuum chamber
(label (1) in Fig.~\ref{fig:setup}). At the endpoints of the translational stage, the direction of motion is reversed and helical motion of the opposite sense occurs. The translational stage (OWIS, model X41.083.101 GP) has a travel range of 95\,mm at a spindle pitch of 0.5\,mm and is
operated at about 200 revolutions per minute. The 520\,mm long connecting shaft transmits
the mechanical motion into the vacuum chamber through a commercial rotary feedthrough
designed for high rotation frequencies and stable guiding of the shaft (K. Lesker) (2).
It is coupled to the spindle of the translation stage and to the sample rod by stiff
couplers. Precise alignment of the axis of the translation stage and the feedthrough is
an important issue since, due to the considerable length of the shaft, a slight
misalignment results immediately in an unbalanced motion at the laser intersection point.
The sample rod can be aligned with respect to nozzle and skimmer using three fine thread
screws connecting two steel plates by ball joints (3). The steel plates are welded onto
the flanges of a bellow, which serves as vacuum lock chamber for rapid exchange of the
sample rods. The translational stage is mounted onto a slider which can be moved along a
four cornered aluminium shaft such that the sample can be pulled out into the bellow
chamber. The whole setup is attached to the source chamber by a manual gate valve (4). In
order to compensate the torque acting onto the flange the four cornered shaft is
supported at its end by a spring.

Previous experiments on laser ablation have shown that clogging of the 10\,$\mu$m-nozzle
and the 0.4\,mm skimmer is a severe drawback in particular when working at high
repetition rates. Various geometries have been tested for shielding nozzle and skimmer
from the vapor expanding out of the ablation plume. Our attempts included sheathing the
sample rod with a metal cylinder with a slit along its axis allowing the laser to
interact with the sample surface. However, no unperturbed ablation conditions were
achieved using this type of shielding. The best results are obtained by placing small sheet metal strips underneath nozzle and skimmer such that the direct line of sight from the laser-surface interaction region to nozzle and skimmer orifices is covered without affecting the helium jet.

The ablation laser is a pulsed frequency-doubled Nd:YLF laser (526\,nm) that produces
pulse energies of up to 15\,mJ at 1\,kHz repetition rate (Quantronix Falcon 263). The
pulse duration is in the range 130 - 250\,ns depending on laser power. A few measurements
for characterizing laser doping at pulse energies exceeding 15\,mJ are performed using a
Nd:YAG laser (1064\,nm) at 10\,Hz repetition rate (Quanta-Ray) with pulse lengths of
about 10\,ns. Although the laser wavelengths and pulse durations are different, no
significant differences in the laser ablation efficiency have been observed.

The laser beam is focused onto the sample rod using a lens of 300\,mm focal length which
is mounted on a $x-$, $y$-, and $z$-translation stage. In vertical direction the lens
position can be varied in a range of 50\,mm in order to change the energy density and the
ablation spot size. The horizontal alignment is important to center the laser beam focus
on the metal rod in between nozzle and skimmer. Following M. Smits and
coworkers~\cite{Smits:2003}, the lens is periodically moved back and forth by 0.5\,mm in
order to avoid the formation of grooves on the sample rod and thus to improve long-term
stability of the ablation signal. This is done by connecting a small DC-motor to the
$x$-translation stage via a flexible shaft ((6) in Fig.~\ref{fig:setup}). Both controllers for this motor and for
the motorized translation stage are home built.

Although the laser entrance window is situated 240\,mm above the ablation sample,
considerable coating of the window surface is an issue when working with metal samples.
Depending on the type of metal used, the coating leads to strong absorption of the laser
light and to noticeable heating of the window. This has even lead to cracking of the
window. For this reason, a thickness of the laser entrance window of 6\,mm or thicker
should be chosen. In order to enable rapid exchange of the window for cleaning or
replacement a manual gate valve is introduced in between the vacuum chamber and the
window mount (5). This deficiency may resolved by shielding the window with a gas buffer
 (see e.g.\ Ref.~\cite{Smits:2003}).

\section{\label{sec:Results}Doping helium nanodroplets by laser ablation}
\subsection{\label{sec:Magnesium}Laser ablation of magnesium}
When using laser ablation as means of doping a helium nanodroplet beam, the desired
process of attaching atoms or ions to the droplets competes with an unwanted effect of
the ablation plume
on the droplet beam: depletion of the beam by inelastic collisions of
the droplets with high-energetic particles at high density.


\begin{figure}
\resizebox{0.9\columnwidth}{!}{\includegraphics {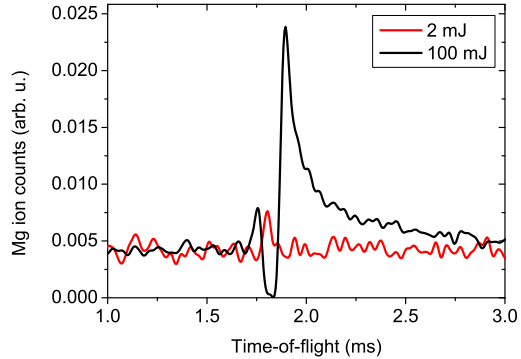}}
\caption{\label{fig:Destruct} Time-of-flight signal traces of helium nanodroplets doped
with Mg for two different laser pulse energies of the Nd:YAG laser operated at 10\,Hz
repetition rate. The signal is measured using a quadrupole mass spectrometer set to a
mass of 24\,amu.}
\end{figure}
The competition between doping and beam destruction is apparent when comparing the
detector signals recorded as a function of delay time after the laser pulse for low
(2\,mJ) and high (100\,mJ) pulse energies, as depicted in Fig.~\ref{fig:Destruct}.
Similar signal traces for ablation of sodium and magnesium have been discussed in detail
in Ref.~\cite{Claas:2003}. The shown signal traces reflect doping by ablation of Mg using
a Nd:YAG laser (1064\,nm) operated at 10\,Hz repetition rate. The laser beam is focused
to a waist of 0.2\,mm. In this experiment a QMS with electron impact ionization at 30\,eV
electron energy is used as a detector. The detector is operated in pulse counting mode
and the detector pulses are discriminated and recorded using a multi-channel scaler. 
The response time of the QMS is in the range 30-40\,$\mu$s which leads to a small
broadening effect on the time structure of the signal.

The ion signal shown in Fig.~\ref{fig:Destruct} reflects the number of ion counts per
10\,$\mu$s time bin and per laser shot measured with our QMS detector. Since the measured count rate strongly
depends on the particular detector used it should by no means be mistaken as an absolute value of
the beam intensity. 
The absolute intensity of doped helium droplets is roughly estimated to be a factor $10^5$ larger
than the measured ion count rate.

The constant signal level of about $5\cdot 10^{-3}$ per bin per laser shot in
Fig.~\ref{fig:Destruct} is attributed to ionization
fragments from residual gas molecules embedded in the helium nanodroplets. At low pulse
energy (2\,mJ), a small signal peak at about 1.6\,ms time of flight is recorded
indicating laser induced doping of the droplets with Mg atoms. At high pulse energy
(100\,mJ), the ion signal is completely depleted in the time interval, in which a
positive doping signal was previously measured. Clearly, the droplet beam is destroyed
due to the interaction with a dense ablation plume. The width of
the signal minimum of about 80\,$\mu$s corresponds to the length of the droplet beam
interval which is exposed to the expanding plasma plume of 15\,mm when taking
time-of-flight broadening into account. However, both at earlier and at later times
doping of the droplet beam prevails. The fact that significant doping takes place up to
1\,ms after the attenuation phase is explained by the ablated particles undergoing
multiple scattering events before being decelerated and sticking to the
droplets~\cite{Claas:2003}. The small positive doping signal at earlier times (1.7\,ms)
might stem from weak doping before a high particle density has built up to deplete the
droplet beam or it might be due to Mg atoms being forward scattered in the direction of
the helium beam through the skimmer and being picked up further downstreams. A clear
signal of forward scattered Mg ions has been observed in earlier experiments in
time-of-flight spectra of ion-doped helium droplets.

\begin{figure}
\resizebox{0.9\columnwidth}{!}{\includegraphics{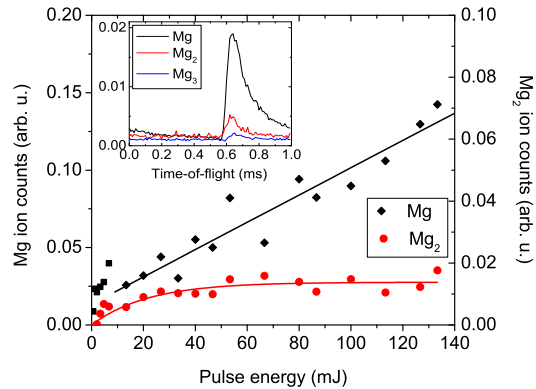}}
\caption{\label{fig:MgTraces} Dependence of the yield of ionized Mg monomers and dimers
as a function of laser pulse energy. Inset: Time-of-flight signal traces of Mg doped
droplets recorded at the mass of the Mg monomer, dimer and trimer.}
\end{figure}
Using the QMS as a detector has the advantage of allowing mass-selective measurements of
the yield of monomers and oligomers of the laser-doped material. Typical time-of-flight
signal traces recorded at the mass of the Mg monomer, dimer and trimer using laser
ablation with the Nd:YLF laser (526\,nm) with a pulse energy of 6.3\,mJ at 1\,kHz
repetition rate are depicted in the inset of Fig.~\ref{fig:MgTraces}. The signal-to-noise
ratio is much better than the one of Fig.~\ref{fig:Destruct} as a result of averaging
over $10^5$ laser shots and due to the better shot-to-shot stability of the pulse energy
of the Nd:YLF laser. Despite of maximum focusing of the laser beam, destruction of the
droplet beam at this pulse energy is barely visible as a small dip of the ion yield right
before the doping signal maxima. Note the falling level of the monomer signal from 0 to
0.6\,ms, which is due to doping by the previous laser shot. Thus, at an ablation rate of
1\,kHz nearly continuous doping of the droplet beam is achieved.


In order to investigate the efficiency of multiple doping at higher pulse energies
similar measurements using the 10\,Hz Nd:YAG laser are carried out.
Fig.~\ref{fig:MgTraces} shows the yield of monomers versus dimers for pulse energies up
to 140\,mJ. The plotted ion count rate is obtained by integrating the count rate over an
interval of width 200\,$\mu$s. 
The curved solid line represents the fit of a simple saturation model and is intended to
guide the eye. Note the different vertical scales for monomer and dimer signals (left and
right, respectively). One observes a local maximum both of monomer and dimer
doping at low energies between 5 and 10\,mJ, similar to doping with Li. This observation will
be discussed in more detail in the following section. At higher pulse
energies, the monomer yield again increases roughly linearly, as indicated by the
straight line. However, the dimer signal saturates at about 50\,mJ at an ion count rate
of 0.015 counts per laser shot and thus always stays well behind the monomer signal.
The same behavior has been observed with all metal samples discussed
below. Thus we conclude that the pick-up process in the laser ablation plume proceeds in
a significantly different way from the one in the vapor inside a heated oven, which is
known to follow Poissonian statistics~\cite{Lewerenz:1995}. Although the relative intensities may
be perturbed by fragmentation processes upon electron impact ionization we don't expect
larger cluster masses to disappear completely~\cite{Tiggesbaeumker:2007}.



\subsection{\label{sec:Lithium}Laser ablation of lithium}
Systematic experiments for characterizing the new laser ablation setup are carried out
using solid Li as sample material for several reasons. Alkali atoms and clusters are very
sensitively detected by a Langmuir-Taylor (LT) surface ionization
detector~\cite{Stienkemeier:2000}. From previous experiments it is known that alkali
metals are efficiently ablated at low laser intensities. Li atoms are easily accessible
by electronic spectroscopy using dye lasers and the spectrum of Li attached to helium
nanodroplets is well known~\cite{Stienkemeier:1996}. Furthermore, Li can be purchased as
metal rods 10\,mm in diameter and can be handled with reasonable safety measures.

\begin{figure}
\resizebox{0.8\columnwidth}{!}{\includegraphics{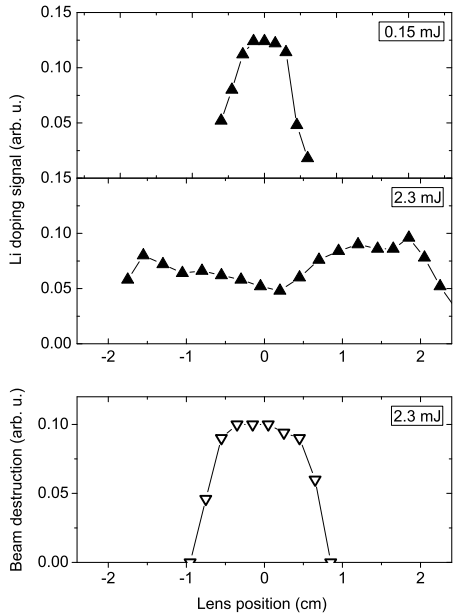}}
\caption{\label{fig:MaximaScan} Amplitude of the doping signal as a function of the
position of the focusing lens for low (0.15\,mJ) (a) and moderate (2.3\,mJ) (b) pulse
energies. Amplitude of the signal drop due to beam destruction at moderate
(2.3\,mJ) pulse energies (c).}
\end{figure}
The measured signal traces are very similar to the ones recorded with Mg, the only
difference being additional broadening due to the response time of the LT-detector of
about 50\,$\mu$s. The amplitude of the signal maxima resulting from laser-doping at two different laser pulse
energies, 0.15\,mJ and 2.3\,mJ, is represented in Fig.~\ref{fig:MaximaScan} (a) and (b),
respectively, as a function of the position of the 300\,mm focusing lens.
The waist of the laser beam in the focus was measured to be $w_0=36\,\mu$m and at 1\,cm
displacement it is $w=160\,\mu$m. At low pulse energy the doping efficiency is clearly
peaked around the focus position~(Fig.~\ref{fig:MaximaScan} (a)). However, at higher
pulse energy maximum doping is achieved when the lens is shifted along the beam by about
1-2\,cm to either direction out of the focus position~(Fig.~\ref{fig:MaximaScan} (b)).
This behavior reflects the fact that for maximum focusing destruction of the beam
partially compensates for higher doping efficiency, whereas outside the focus position,
where energy density is low, only doping occurs. Thus, two distinct regimes of laser-matter
interaction appear to be active: At low energy density, vaporization of material into the
gas phase with low kinetic energy transfer occurs. In this regime, a luminescent plasma
at the laser -- rod interaction point is barely visible. At higher energy density,in
contrast, a bright plasma plume can be seen. In this regime, the ejected material
acquires large kinetic energy, presumably due to more violent interactions inside the
strongly ionized plasma.

Fig.~\ref{fig:MaximaScan} (c) displays the beam depletion due to the laser ablated
material as a function of lens position at 2.3\,mJ pulse energy. Clearly, the maximum
signal drop is correlated with the local minimum of the doping signal shown in
Fig.~\ref{fig:MaximaScan} (b). The steep edges of both the doping maximum in
Fig.~\ref{fig:MaximaScan} (a) and the depletion maximum in Fig.~\ref{fig:MaximaScan} (c)
indicate that both processes exhibit a threshold behavior as a function of laser energy,
however, with different threshold values.

\begin{figure}
\resizebox{0.9\columnwidth}{!}{\includegraphics{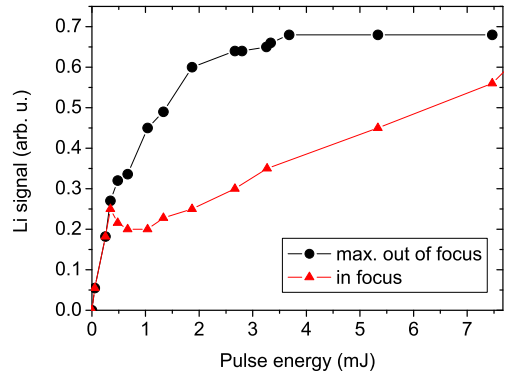}}
\caption{\label{fig:PulseE} Amplitudes of the laser doping signal as a function of laser
pulse energy at the lens position of the maximum doping by desorption (circles) and in
focus (triangles).}
\end{figure}
Fig.~\ref{fig:PulseE} depicts the doping signal amplitudes at the focus position
(triangles) and at the position of maximum doping efficiency outside the focus (dots).
Strikingly, at about 0.3\,mJ pulse energy beam destruction suddenly sets in for the case
of tight focusing. With increasing pulse energy, the doping signal even decreases to a
local minimum at about 1\,mJ before rising again almost linearly. As mentioned in the
previous section, a similar behaviour occurs when ablating Mg.  The doping signal
continuously increases even when the laser pulse energy goes up to 200\,mJ, as observed
using the 10\,Hz Nd:YAG laser. Thus, the density of particles amenable to pick-up by
helium droplets by scattering with helium gas further increases. In this regime of laser
ablation, large amounts of material are ejected into the gas phase which can be seen from
the fast rate of removal of material from the sample and from coating of the laser
entrance window. The doping amplitude at the lens position of most efficient doping
outside the focus (vaporization), however, continuously increases before saturating. From
analyzing the dependence of peak position and pulse energy we find that maximum doping by
vaporization is correlated with a constant energy density per pulse of 50 J/cm$^2$ in the
case of ablating Li, which is nearly independent of the diameter of the laser spot on the
sample surface. The initial signal rise is then explained by a growing effective surface
area from which the atoms are vaporized. As the laser waist on the surface reaches
0.3\,mm at about 3\,mJ pulse energy, the density of vaporized slow Li atoms becomes
sufficiently high that the droplet beam is attenuated by multiple pick-up of atoms.
Besides, at this size of the laser spot the finite curvature of the sample surface is
expected to play a role. Thus, for laser doping using moderate pulse energies in the
range of 0.5-10\,mJ, which are typical for kHz lasers, optimal conditions are obtained
when defocusing the laser beam on the sample to a waist of 0.1-0.5\,mm.

\begin{figure}
\resizebox{0.9\columnwidth}{!}{\includegraphics{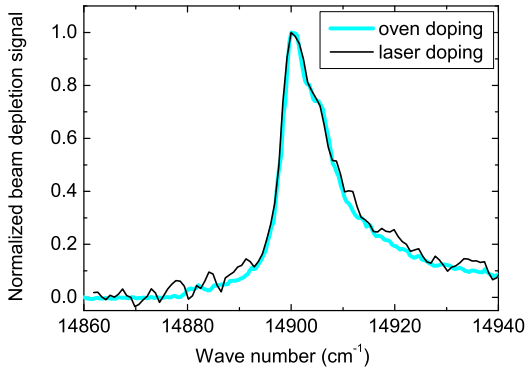}} \caption{\label{fig:LiSpec}
Spectrum of helium nanodroplets doped with Li atoms by means of laser ablation (red line)
in comparison with oven doping.}
\end{figure}
The objective of implementing a kHz laser ablation source into our apparatus is to carry
out electronic spectroscopy of atoms, molecules, and ions inside helium droplets, which
are not amenable to doping by evaporation in an oven. As a first demonstration
experiment, the beam depletion spectrum of the D2 atomic transition of Li attached to
helium droplets is measured with laser doped Li and, for comparison, with Li doped using
a conventional oven. The resulting spectra are depicted in Fig.~\ref{fig:LiSpec}. In the
experiment, the beam of a cw dye laser is superimposed coaxially with the helium droplet
beam ($\approx50$\,mW laser power coupled into the machine). The depletion of the doped
helium droplet beam upon electronic excitation is measured with the LT-detector. The two
spectra nicely match. However, the signal-to-noise ratio for ablation-doped Li is much
lower with respect to oven-doping. One reason is the given lower duty-cycle of about 10\%
for laser doping. Secondly, since the laser doping has not been optimized for picking up
one atom per droplet only, the LT detector signals also have contributions from lithium
oligomers which do not show up in the observed depletion spectrum. Finally, shot to shot
fluctuations diminish signal to noise ratios. In our case Li samples do not have a
perfectly flat surface since we are dealing with a soft material that has to be cleaned
by scraping off the oxidized surface layer with a knife prior to mounting onto the sample
coupler. Consequently, for the harder metal samples discussed below, shot-to-shot
fluctuations are found to be smaller although the absolute signal rates are lower.

\subsection{\label{sec:Refr}Doping with non-volatile metals}

\begin{figure}
\resizebox{0.9\columnwidth}{!}{\includegraphics{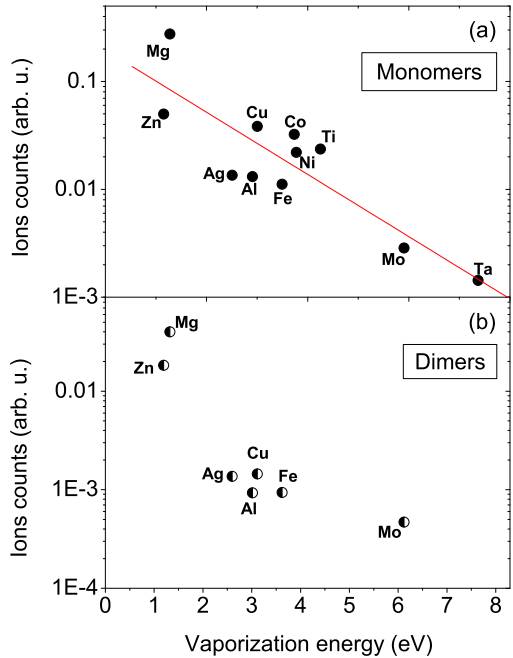}}
\caption{\label{fig:VapEnergy} Yield of various metal monomers (a) and dimers (b)
detected by the QMS as a function of vaporization energy \cite{PeriodicTable}.}
\end{figure}

In order to test the new ablation setup for doping helium nanodroplets with high
temperature refractory materials, a variety of metal samples have been tried out. On the
one hand, the ability of doping helium nanodroplets with pure metal atoms is of interest
for extending the panoply of systems for spectroscopic studies. Moreover, compounds of
organic molecules and metal atoms are of considerable interest for potential applications
as catalysts. On the other hand, the isolation of metal clusters inside helium droplets
at a temperature of 0.4\,K would open the way to studying the magnetic and
superconducting properties of such nano-sized systems.

Fig.~\ref{fig:VapEnergy} gives an overview over the metals studied so far. The laser is
operated at maximum pulse energy of about 10\,mJ and the beam focus is placed onto the
sample surface. The yield of ionized monomers and dimers detected by the QMS is
represented in a logarithmic scale as a function of the vaporization energy in
Fig.~\ref{fig:VapEnergy} (a) and (b), respectively. The plotted ion counts reflects the
average number of detected ions per laser shot within a time window of 200\,$\mu$s. The
different detection efficiencies for the different masses are not corrected for. The straight line in
Fig.~\ref{fig:VapEnergy} (a) represents the fit of an exponentially decreasing function
to indicate the general trend as observed with these selected metals. The efficiency of
doping helium droplets with tantalum is more than 2 orders of magnitude lower than for
doping with Mg. Fig.~\ref{fig:VapEnergy} (b) shows only those materials for which a
significant dimer signal was measured. This is not the case for copper, cobalt, nickel,
titanium, and tantalum. Obviously, the ablation efficiency does not merely depend on the
vaporization energy, but is related in a complex fashion to a number of parameters
including heat capacity, heat conductivity, reflectivity of the surface, absorption
spectrum of ablated atoms, etc. The important point here is that all materials tested so
far can be laser ablated and lead to a measurable doping signal. No significant formation
yield of metal clusters has been achieved in our experiments. However, parameters like
e.g. the droplet size distributions, which has been shown to influence significantly
multiple pickup, has not been optimized for metal cluster formation so far.


\begin{figure}
\resizebox{0.85\columnwidth}{!}{\includegraphics{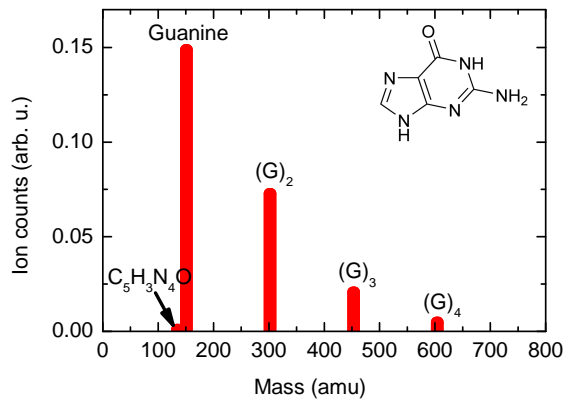}}
\caption{\label{fig:Guanin} Yield of guanine and guanine oligomers doped in helium nanodroplets
by laser ablation.}
\end{figure}
\subsection{\label{sec:Biomol}Doping with biomolecules}
Doping helium nanodroplets with biomolecules is of considerable interest in the context
of matrix isolation spectroscopy and photobiology at very low
temperatures~\cite{Zewail:2000}. Thermalization of large molecules to the equilibrium
temperature of helium nanodroplets of 0.4\,K is known to tremendously simplify the
structure of the spectra with only negligible perturbation due to the interaction of the
molecule with the helium environment~\cite{Lindinger:1999,Huisken:1999}. The neat DNA
base guanine is used as the first demonstration system. As for many biomolecules, simple
heating would lead to thermal decomposition before vaporization~\cite{Piuzzi:2000}. Laser
desorption of guanine at high repetition rate was also demonstrated by Smits \textit{et
al.}~\cite{Smits:2003}. 

The sample is prepared by pressing neat guanine into cylindrical
shape onto a threaded rod. Fig.~\ref{fig:Guanin} displays the yield of guanine
monomers and oligomers as detected by the QMS. The laser pulse energy is set to 1.5\,mJ
at maximum focusing of the laser beam. Under these conditions, laser desorption at 1\,kHz
repetition rate is stable for more than one hour. The count rate of the monomer signal is
found to be comparable to the one achieved by Smits \textit{et al.} using femtosecond
resonant photoionization. Only negligible fragmentation is observed, the largest fragment
contribution being C$_5$H$_3$N$_4$O at a mass of 135\,amu with only 0.7\,\% of monomer signal. The relative
yields of dimers, trimers and tetramers is found to be slightly larger compared to the
observations of Smits \textit{et al.}. This is quite surprising given the fact that
clustering of metal atoms was found to be much less efficient in our experiment but may
be due to the larger droplet sizes used. 

With laser desorption of guanine, the
maximum signal yield is obtained for a nozzle temperature of 14\,K and a stagnation
pressure of 54\,bar which corresponds to expansion conditions close to the transition to
the expansion out of liquid helium (supercritical expansion). At these source conditions,
the average number of helium atoms per droplet is expected to lie in the range 1-2$\cdot 10^4$. The need for
large helium droplets is attributed to the fact that large molecules such as guanine
carry large amounts of internal energy due to their large number of internal degrees of
freedom. Consequently, cooling to 0.4\,K requires the ability of the droplets to
evaporate a large number of helium atoms.

\begin{figure}
\resizebox{1.0\columnwidth}{!}{\includegraphics{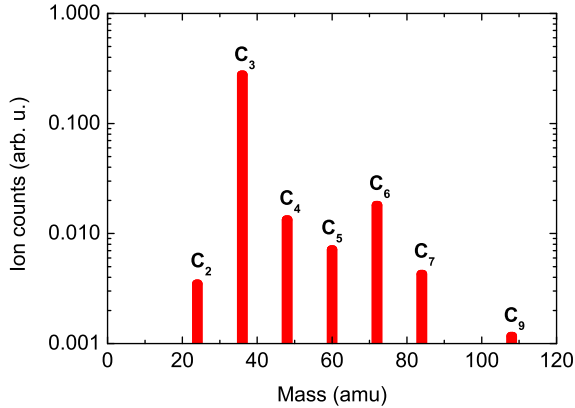}}
\caption{\label{fig:Graphit} Yield of small carbon clusters doped in helium nanodroplets by
ablation of a graphite sample.}
\end{figure}
\subsection{\label{sec:Carbon}Laser evaporation of carbon samples}
Besides pressing samples of neat material, large organic molecules are commonly prepared
in samples consisting of the molecules and a matrix material. Since the standard matrix
material for laser desorption is graphite we chose neat graphite as another material for
testing our setup. 

Typical QMS count rates measured at different oligomer masses are
depicted in Fig.~\ref{fig:Graphit}.
The monomer signal is not shown because of a large background count rate at the mass of
12\,amu which stems from the ionization of hydrocarbons in the residual gas of the vacuum
system and from He$_3^+$ fragments. An upper limit for the monomer signal rate can be estimated to 0.05 events per
laser shot. C$_3^+$ is by far the dominating cluster size with a count rate of
0.3\,s$^{-1}$. This is the highest count rate of all materials studied so far. This observation is
consistent with earlier measurements on the fragmentation of small carbon clusters~\cite{Rohlfing:1984,Ramanathan:1993}.
It is due to the local stability of C$_3$ and C$_3^+$ clusters~\cite{Raghavachari:1987}. As for
doping with guanine, the highest yield of carbon oligomers is obtained at considerably
lower nozzle temperatures (19\,K for C$_2$, 18\,K for C$_3$, 15\,K for C$_4$, and 14\,K
for C$_5$, C$_6$, C$_7$, C$_8$). 

In this case we interpret the need for large helium
droplets by the large amount of binding energy (6.4, 13.9, 19.0, 26.3, 31.8, 38.7, 44.3, 51.0\,eV for neutral clusters C$_n$, where $n$ runs from 2 to 9, respectively) liberated upon cluster formation during
the interaction of carbon atoms with helium~\cite{Raghavachari:1987}. Despite the high signal yield obtained at maximum laser power and tight
focusing, the sample surface is only weakly degraded after more than 2 hours of laser
ablation at 1\,kHz repetition rate. The laser entrance window remained nearly uncoated,
such that the laser-doping signal rate was stable for many hours of operation.

\section{\label{sec:Summary}Summary}
In conclusion we demonstrated that helium nanodroplets can be doped quite efficiently by
means of laser ablation at kHz repetition rates. The method is characterized in terms of
laser fluences, focussing conditions, etc. For the first time, metals have been doped for
which thermal evaporation is not an option. The intensities of doping the different
materials are compared quantitatively. Extending the doping method to biomolecules,
attachement of guanine to helium droplets has been demonstrated. The unfragmented
monomers as well as oligomers have been observed. Finally, graphite samples provide as
all the other materials stable evaporation conditions. In combination with other organic
samples this opens the possibility to use matrix assisted laser desorption to extend
studies at ultracold temperatures in helium droplets to a variety of  molecules of
biological interest.

\section{\label{sec:Acknowledgements}Acknowledgements}
The authors gratefully acknowledge stimulating discussions with A. Stolow and J.
Tiggesb\"aumker and financial support by the Deutsche Forschungsgemeinschaft.

\newpage 

\end{document}